\documentclass[submission,copyright,creativecommons]{eptcs}
\providecommand{\event}{MCU 2013} 
\usepackage{breakurl}             
\usepackage{epsfig}
\usepackage{color}

\title{Reversible Logic Elements with Memory and\\ Their Universality}
\author{Kenichi Morita%
\thanks{Currently, Professor Emeritus of Hiroshima University}
\institute{Hiroshima University, Higashi-Hiroshima, 739-8527, Japan}
\email{km@hiroshima-u.ac.jp}
}
\def\titlerunning{Reversible Logic Elements with Memory}
\def\authorrunning{K. Morita}

\newtheorem{thm}{Theorem}

\newtheorem{lem}{Lemma}

\newtheorem{defn}{Definition}

\definecolor{darkred}{rgb}{0.8,0,0}

\begin{document}
\maketitle

\begin{abstract}
Reversible computing is a paradigm of computation that 
reflects physical reversibility, one of the 
fundamental microscopic laws of Nature. 
In this survey, we discuss topics on reversible logic 
elements with memory (RLEM), which can be used to build 
reversible computing systems, and their universality. 
An RLEM is called universal, if any reversible sequential 
machine (RSM) can be realized as a circuit composed only of it. 
Since a finite-state control and a tape cell of a reversible 
Turing machine (RTM) are formalized as RSMs, any RTM can 
be constructed from a universal RLEM. 
Here, we investigate 2-state RLEMs, and show that infinitely 
many kinds of non-degenerate RLEMs are ``all" universal 
besides only four exceptions. 
Non-universality of these exceptional RLEMs is also argued. 
\end{abstract}


\section{Introduction}

A reversible computing system is a one such that every 
computational configuration of it has at most one 
predecessor, i.e., a ``backward deterministic" system. 
Though the definition is thus simple, it is known that 
it has a close relation to physical reversibility. 
Since physical reversibility is one of the fundamental 
microscopic laws of Nature, it is important how this 
property is utilized to construct an efficient 
reversible computers. 
So far, many kinds of reversible computing models have 
been proposed and investigated. 
We should note that there are several levels of models 
ranging from a microscopic one to a macroscopic one. 
In the bottom (i.e., the most microscopic) level, there is 
a physically reversible model, e.g., the billiard ball 
model (BBM) of computing \cite{FT82}. 
In the next level, there exist various kinds of 
reversible logic elements such as Fredkin gate \cite{FT82}, 
Toffoli gate \cite{Tof80,Tof81}, and reversible logic 
elements with memory \cite{Mor01R}. 
In the still higher level, there are reversible logic 
circuits composed of reversible logic elements, which can 
be  used as building modules for reversible computers. 
In the top level, there are models of reversible computers 
such as reversible Turing machines \cite{Ben73}, 
reversible cellular automata \cite{Tof77}, and others. 

Here, we focus on the topics of a reversible logic element. 
It is a primitive for composing reversible logic circuits 
whose function is described by a one-to-one mapping. 
There are two types of such elements: 
one without memory, which is usually called a reversible 
logic gate, and one with memory. 
The conventional design theory of logic circuits has been 
developed using logic gates as primitives (but in the study 
of asynchronous circuits, logic elements with memory are 
sometimes used \cite{BP80,Kel74}).
On the other hand, in the case of reversible computing, 
logic elements with memory are also useful. 
The main reason is that if we use an appropriate reversible 
logic element with memory, we can construct several kinds of 
reversible computing models, e.g., reversible Turing machines, 
very simply \cite{Mor01R,Mor10R}.  

In this paper, we give a survey on {\em reversible logic 
elements with memory} (RLEM) based mainly on the studies of 
the author and his colleagues. 
In particular, we focus on the topics of universality of RLEMs. 
An RLEM is called universal, if any reversible sequential 
machine (RSM) can be realized by it. 
Since a reversible Turing machine (RTM), which is  
a universal computing model \cite{Ben73}, is composed of RSMs, 
we can construct any RTM using a universal RLEM. 
Here, we investigate 2-state RLEMs, i.e., RLEMs with 
1-bit memory.  
There are infinitely many 2-state RLEMs if we do 
not restrict the numbers of input/output symbols. 
We shall see that ``all" the non-degenerate 2-state 
RLEMs except only four 2-symbol RLEMs are universal. 
We also discuss non-universality of these 2-symbol RLEMs. 


\section{Reversible logic element with memory (RLEM)}

We first give a definition of sequential machine (SM), 
since a reversible logic element with memory (RLEM) is 
a special type of an SM. 
An SM considered here is a kind of a finite automaton 
with an output port as well as an input port, which is 
often called an SM of Mealy type. 

\begin{defn}\label{DEF:RSM}
A {\em sequential machine} (SM) is a system defined by 
$M = (Q, {\Sigma}, {\Gamma}, \delta)$, 
where $Q$ is a finite set of internal states, 
${\Sigma}$ and ${\Gamma}$ are finite sets of 
input and output symbols, and 
$\delta: Q\times{\Sigma}\rightarrow Q\times{\Gamma}$ 
is a move function. 
If $\delta$ is injective, $M$ is called 
a {\em reversible sequential machine} (RSM).
Note that if $M$ is reversible, then 
$|{\Sigma}| \le |{\Gamma}|$ must hold. 
A {\em reversible logic elements with memory} (RLEM) is 
an RSM $M = (Q, {\Sigma}, {\Gamma}, \delta)$ such that 
$|{\Sigma}| = |{\Gamma}|$. 
In particular, it is called a $|Q|$-state 
$|{\Sigma}|$-symbol RLEM. 
\end{defn}

\newcommand{\rmH}{%
 \put(-0.8, -0.8){\thicklines\framebox(1.6,1.6){}}
 \put(-0.7, -0.0){\line(1,0){1.4}}
 \put( 0.0,  0.0){\circle*{0.30}}
}
\newcommand{\rmV}{%
 \put(-0.8, -0.8){\thicklines\framebox(1.6,1.6){}}
 \put( 0.0,  0.7){\line(0,-1){1.4}}
 \put( 0.0,  0.0){\circle*{0.30}}
}
\newcommand{\rmHc}{%
\begin{picture}(1.8,1.8)(-0.9,-0.9)
\put(0,-0.3){\rmH}
\end{picture}
}
\newcommand{\rmVc}{%
\begin{picture}(1.8,1.8)(-0.9,-0.9)
\put(0,-0.3){\rmV}
\end{picture}
}
\setlength{\unitlength}{0.5em}

\begin{figure}[h]
\begin{center}
 \includegraphics[scale=0.7]{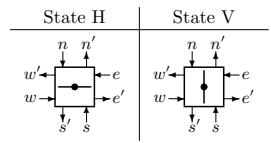}
 \caption{Two states of a rotary element (RE). 
 \label{FIG:re_two_states} 
 }
\end{center}
\end{figure}

\begin{figure}[h]
\begin{center}
 \includegraphics[scale=0.7]{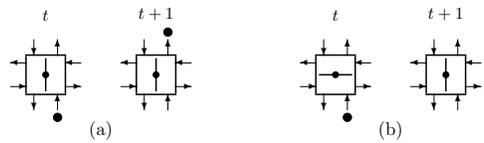}
 \caption{Operations of a rotary element (RE): (a) the parallel case, and 
 (b) the orthogonal case. 
 \label{FIG:REoperation} 
 }
\end{center}
\end{figure}

Hereafter, we mainly discuss 2-state RLEMs. 
There are infinitely many kinds of RLEMs if we do not 
limit the number of symbols. 
Among them, a {\em rotary element} (RE) \cite{Mor01R} is 
a typical RLEM with four symbols. 
Its behavior can be very easily understood, since 
it has the following interpretation on its operation. 
An RE is depicted by a box that contains a rotatable bar 
inside (Fig.~\ref{FIG:re_two_states}). 
Two states of an RE are distinguished by the direction 
of the bar, and thus they are called state H and state V. 
There are four input lines and four output lines 
corresponding to the sets of input symbols $\{n,e,s,w\}$ 
and output symbols $\{n',e',s',w'\}$. 
The rotatable bar is used to control the move direction 
of an input signal (or a particle). 
When no particle exists, nothing happens on the RE. 
If a particle comes from the direction parallel to the 
rotatable bar, then it goes out from the output line of 
the opposite side  without affecting the direction of 
the bar (Fig.~\ref{FIG:REoperation}~(a)). 
If a particle comes from the direction orthogonal to the 
bar, then it makes a right turn, and rotates the bar by 
90 degrees counterclockwise (Fig.~\ref{FIG:REoperation}~(b)). 
It is reversible in the following sense: 
from the next state and the output, the previous state 
and the input are uniquely determined.  
Actually, an RE is defined as the following RSM: 
$M_{\rm RE} = (\{\,\rmHc\,,\,\rmVc\,\},
\{n,e,s,w\},\{n',e',s',w'\},\delta_{\rm RE})$, 
where $\delta_{\rm RE}$ is given in Table~\ref{TABLE:table_re}.

\begin{table}[ht]
\begin{center}
 \caption{The move function $\delta_{\rm RE}$ of a rotary element (RE). 
 \label{TABLE:table_re} 
 }
{\small
\setlength{\unitlength}{0.5em}
\begin{tabular}{c||c|c|c|c|}
                &   \multicolumn{4}{c|}{Input}   \\\cline{2-5}
{Present state} &     $n$      &     $e$      &     $s$      &     $w$      \\\hline\hline
{State H:}\ \rmHc &\ \rmVc $w'$\ &\ \rmHc $w'$\ &\ \rmVc\,$e'$\ &\ \rmHc $e'$\ \\\hline
{State V:}\ \rmVc &\ \rmVc\ $s'$\ &\ \rmHc\,$n'$\ &\ \rmVc $n'$\ &\ \rmHc $s'$\ \\\hline
\end{tabular}
}
\end{center}
\end{table}

\begin{figure}[h]
\begin{center}
 \includegraphics[scale=0.49]{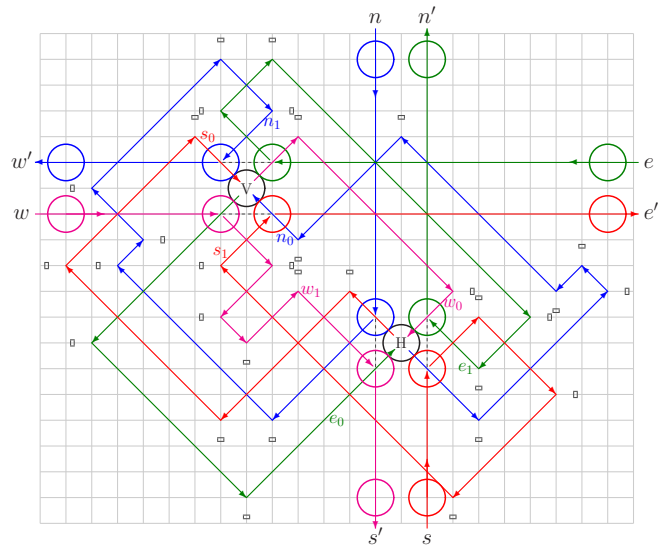}
 \caption{\label{FIG:bbm_re_all}
 A rotary element (RE) realized in BBM \cite{Mor08,Mor12}. 
 }
\end{center}
\end{figure}

Now, we consider how reversible logic elements 
can be realized in a reversible physical system. 
In our present technology, it is difficult to implement 
a reversible logic element in a practical system having 
physical reversibility in nano-scale level. 
However, some thought experiments in an idealized 
circumstance suggest a possibility of realizing it. 
The {\em billiard ball model} (BBM) is a reversible physical 
model of computing proposed by Fredkin and Toffoli~\cite{FT82}. 
It is an idealized mechanical model consisting of balls and 
reflectors. 
They showed a Fredkin gate is realizable in BBM. 
On the other hand, an RE can be simulated in BBM as shown 
in Fig.~\ref{FIG:bbm_re_all}~\cite{Mor08,Mor12}. 
It consists of one stationary ball called a state ball, 
and many reflectors indicated by small rectangles. 
A state ball is placed at the position of H or V 
in Fig.~\ref{FIG:bbm_re_all} depending on the 
state of the simulated RE. 
A moving ball called a signal ball can be given 
to any one of the input lines $n, e, s$, and $w$. 
Then, the operation of an RE is correctly simulated by 
collisions of balls and reflectors 
(the details of the movements are found in \cite{Mor10R}). 
In \cite{MM12}, it is shown that any $m$-state $k$-symbol 
RLEM can be realized in BBM in a systematic way when $k \le 4$. 

\section{Constructing reversible machines by RLEMs}

We define universality of an RLEM as the property that any 
RSM can be composed of it, and show an RE is universal. 
We then explain that a reversible Turing machine can also be
constructed using REs. 

\begin{defn}
An RLEM is called {\em universal} if any RSM is realized 
by a circuit composed only of copies of the RLEM. 
\end{defn}

\begin{figure}[h]
\begin{center}
 \includegraphics[scale=0.7]{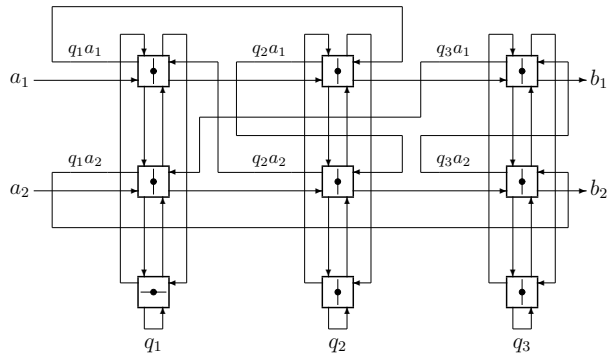}
 \caption{\label{FIG:RE_rsm}
 The RSM $M_0$ implemented by REs \cite{Mor03R}. 
 Here, $M_0$ is in the state $q_1$ since the bottom 
 RE of the leftmost column is in the state H. 
 }
\end{center}
\end{figure}

We can see that any RSM can be realized by a circuit 
composed only of REs~\cite{Mor03R}. 
We explain it by an example. 
Consider an RSM 
$M_0 = (\{q_1,q_2,q_3\},\{a_1,a_2\},\{b_1,b_2\},\delta_0\}$, 
where $\delta_0$ is given in Table~\ref{TABLE:table_rsm}. 
Then, we can construct a circuit composed only of REs that 
simulate $M_0$ as shown in Fig.~\ref{FIG:RE_rsm}. 
Note that when constructing a reversible circuit, 
fan-out of an output is not allowed, and the circuit in 
Fig.~\ref{FIG:RE_rsm} satisfies it. 
The circuit has three columns of REs, each of which 
corresponds to a state of $M_0$. 
If $M_0$'s state is $q_j$, then the bottom RE of the 
$j$-th column is set to the state H. 
All other REs are set to V. 
The REs of the $i$-th row corresponds to the input 
symbol $a_i$ as well as the output symbol $b_i$. 
In Fig.~\ref{FIG:RE_rsm}, the circuit is in the state $q_1$. 
If a particle is given to the line e.g. $a_2$, then after 
setting the bottom RE of the 1st column to V, the particle 
appears on the line $q_1 a_2$, i.e., the crossing point of 
the 2nd row and the 1st column is found.  
Since $\delta(q_1,q_2) = (q_3,b_2)$, this line is 
connected to the RE of the 2nd row of the 3rd column. 
By this, the bottom RE of the 3rd column is set to H, and finally the 
particle appears on the output line $b_2$. 
By generalizing the above construction method, 
we see that any RSM can be realized by REs, and 
thus we obtain the following theorem (its precise 
proof is omitted here).  

\begin{thm}\label{THM:RE_universality} 
{\rm \cite{Mor03R}}\ \ 
A rotary element (RE) is universal. 
\end{thm}

\begin{table}[h]
\begin{center}
 \caption{\label{TABLE:table_rsm}
 The move function $\delta_0$ of an example of an RSM $M_0$. 
 }
{\small
\begin{tabular}[b]{c||c|c|}
              & \multicolumn{2}{c|}{Input} \\\cline{2-3}
Present state &   $a_1$   &   $a_2$    \\\hline\hline
$q_1$         & $q_2 b_1$ & $q_3 b_2$  \\\hline
$q_2$         & $q_2 b_2$ & $q_1 b_1$  \\\hline
$q_3$         & $q_1 b_2$ & $q_3 b_1$  \\\hline
\end{tabular}
}
\end{center}
\end{table}

A reversible Turing machine (RTM) is a TM having backward 
deterministic property (see, e.g., \cite{Ben73,Mor08} 
for its definition). 
It is known that for any irreversible TM, there is an RTM that 
simulates the former and leaves no garbage information when it 
halts \cite{Ben73}, hence RTMs are computationally universal. 
We can see that any RTM can be constructed using only REs 
relatively easily, since a finite-state control and a tape 
cell of an RTM can be formalized as RSMs \cite{Mor01R,Mor10R}.

\begin{figure}[h]
\begin{center}
 \includegraphics[scale=0.78]{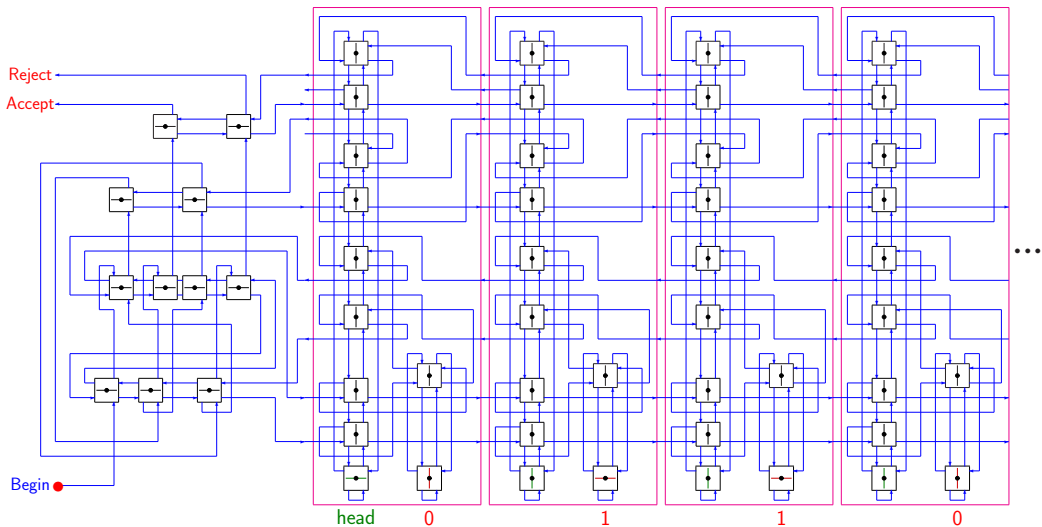}
 \caption{ \label{FIG:re_rtm_parity} 
 A circuit made of REs that simulates an RTM 
 $T_{\rm parity}$ that accepts $\{1^{2n}|\,n=0,1,\ldots\}$. 
 An example of its whole computing process is 
 shown in 4406 figures in \cite{Mor10R}.
 }
\end{center}
\end{figure}

Fig.~\ref{FIG:re_rtm_parity} is a circuit that simulates 
an RTM $T_{\rm parity}$ that accepts the language 
$\{1^{2n}|\,n=0,1,\ldots\}$, whose move function is specified 
by the following set of quintuples: 
\[\{
{[\ } q_0, 0, 1, R, q_1 {\ ]}, 
{[\ } q_1, 0, 1, L, q_{\rm acc} {\ ]}, 
{[\ } q_1, 1, 0, R, q_2 {\ ]}, 
{[\ } q_2, 0, 1, L, q_{\rm rej} {\ ]}, 
{[\ } q_2, 1, 0, R, q_1 {\ ]} \} 
\]
If we give a signal (or a particle) to the input port ``Begin," 
then it starts to compute. 
Finally, the particle comes out from the output port 
``Accept" or ``Reject" depending on the input. 
Detailed descriptions of this circuit as well as how it 
works are given in \cite{Mor10R,Mor12}.

\section{All non-degenerate 2-state RLEMs but four are universal}

In the previous section we saw that an RE is universal. 
On the other hand, since there are infinitely many RLEMs, 
there will be many other universal RLEMs. 
Surprisingly, non-degenerate 2-state RLEMs except only four 
are all universal \cite{MOAT12}. 
In this section, we explain how it is shown. 

\begin{figure}[t]
\begin{center}
 \includegraphics[scale=0.8]{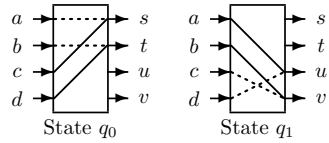}
 \caption{A pictorial representation of the 2-state 
   RLEM 4-289, which is equivalent to RE. 
 \label{FIG:RLEM4-289a} 
 }
\end{center}
\end{figure}

First, we classify 2-state RLEMs. 
We can see the total number of 2-state $k$-symbol RLEMs is $(2k)!$, 
and they are numbered from $0$ to $(2k)!-1$ in some 
lexicographic order \cite{MOTK05R}. 
To indicate that it is a $k$-symbol RLEM, the prefix 
\mbox{``$k$-"} is attached to its serial number like RLEM 4-289. 
Here, we use a pictorial representation of a 2-state RLEM. 
Consider, as an example, a 2-state 4-symbol RLEM 4-289 with the 
input alphabet $\{a,b,c,d\}$, the output alphabet $\{s,t,u,v\}$, 
and the move function given in Table~\ref{TABLE:table_289}. 
Then, it is represented by Fig.~\ref{FIG:RLEM4-289a}, 
where solid and dotted lines in a box describe the 
input-output relation for each state. 
A solid line shows the state goes to another, 
and a dotted line shows the state remains unchanged. 
For example, if the RLEM 4-289 receives an input symbol $c$ 
in the state $q_0$, then it gives the output $s$ and 
enters the state $q_1$. 
As in the case of an RE, we interpret that 
each input/output symbol represents an occurrence of a 
signal at the corresponding input/output port.

\begin{table}[h]
\begin{center}
 \caption{The move function of the 2-state RLEM 4-289. 
 \label{TABLE:table_289} 
 }
{\small
\begin{tabular}{c||c|c|c|c|}
              &   \multicolumn{4}{c|}{Input}   \\\cline{2-5}
Present state &    $a$     &    $b$     &    $c$     &    $d$  \\\hline\hline
 State $q_0$  &\ $q_0\,s$\ &\ $q_0\,t$\ &\ $q_1\,s$\ &\ $q_1\,t$\ \\\hline
 State $q_1$  &\ $q_0\,u$\ &\ $q_0\,v$\ &\ $q_1\,v$\ &\ $q_1\,u$\ \\\hline
\end{tabular}
}
\end{center}
\end{table}

\newpage

\begin{figure}[h]
\begin{center}
\includegraphics[scale=0.9]{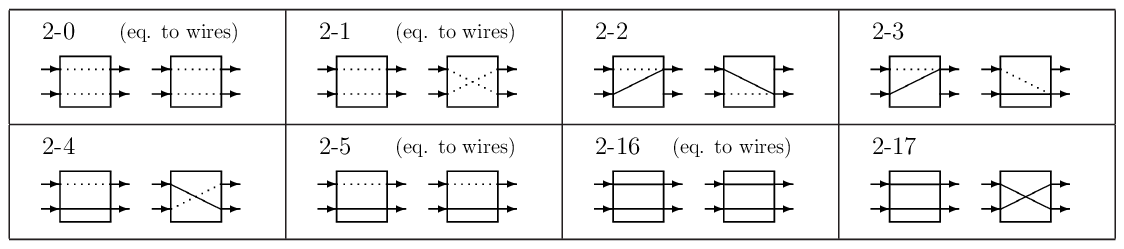}%
\medskip

\includegraphics[scale=0.9]{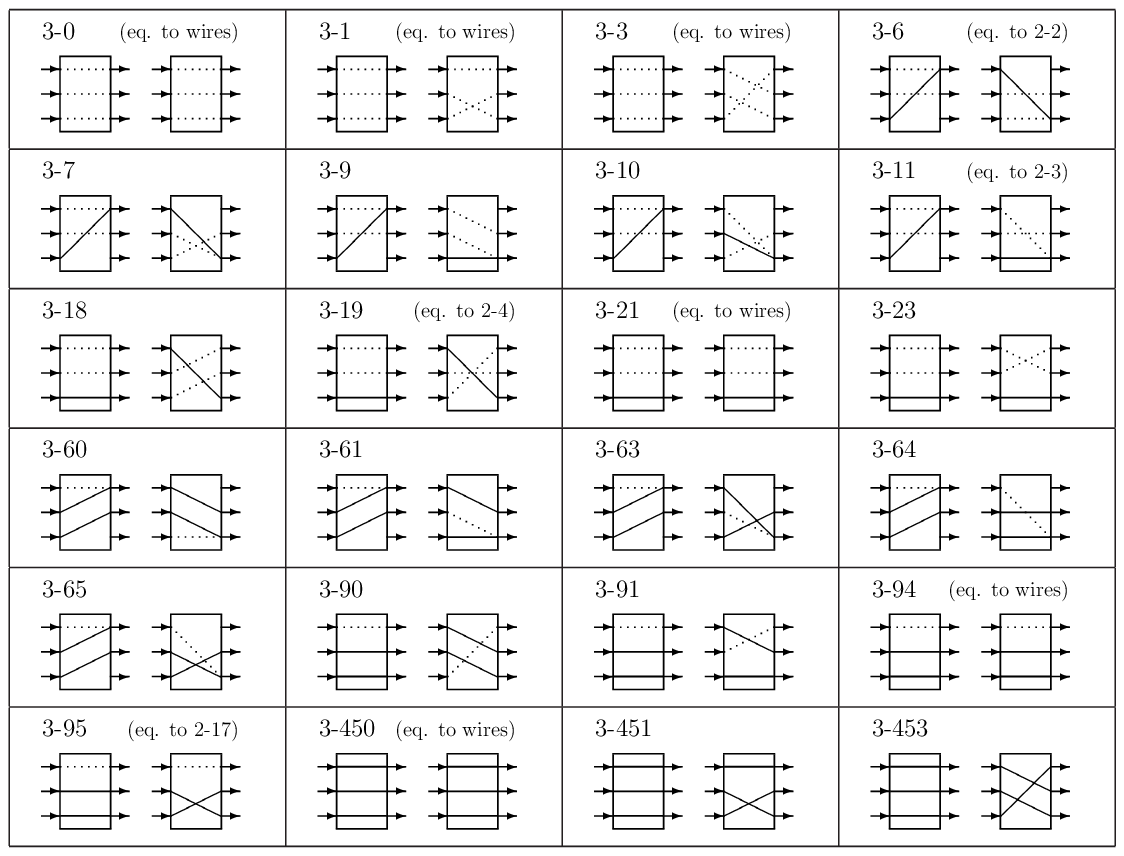}%
\caption{%
   Representatives of 
   8 equivalence classes of 24 2-symbol RLEMs (top), and those of 
   24 equivalence classes of 720 3-symbol RLEMs (bottom) \cite{MOTK05R}. 
   The indications ``eq.~to wires" and ``eq.~to 2-$n$" 
   mean it is equivalent to connecting wires, and 
   it is equivalent to RLEM 2-$n$, respectively. 
   Thus they are degenerate ones. 
   The numbers of 2- and 3-symbol non-degenerate RLEMs are 4 and 14, respectively. 
 \label{FIG:rlem2and3} 
 }
\end{center}
\end{figure}

We can regard two RLEMs are {\em equivalent} if one can be 
obtained by renaming the states and/or the input/output 
symbols of the other. 
It has been shown that the numbers of equivalence classes 
of 2-state 2-, 3-, and 4-symbol RLEMs are 8, 24, and 82, 
respectively \cite{MOTK05R}. 
Fig.~\ref{FIG:rlem2and3} shows all 
representative RLEMs in the equivalence class of 
2- and 3-symbol RLEMs. 
The representatives are so chosen that it has the 
smallest number in the class. 

Among $k$-symbol RLEMs, there are {\em degenerate} ones, 
each of which is either equivalent to simple connecting wires 
(e.g., RLEM 3-3), 
or equivalent to a $k'$-symbol RLEM such that $k' < k$ 
(e.g., RLEM 3-6).  
Its precise definition is found in \cite{MOAT12}. 
In Fig.~\ref{FIG:rlem2and3}, they are 
indicated by ``eq.~to wires" or ``eq.~to 2-$n$". 
Thus, {\em non-degenerate} $k$-symbol RLEMs are the 
main concern of the study. 
It is known that the numbers of non-degenerate 
2- 3- and 4-symbol RLEMs are 4, 14, and 55, respectively.

It has been shown that the following three lemmas hold. 
\begin{lem}\label{LEM:RE_by_3-10}{\rm \cite{LPAM08R,MOAT12}}\ 
An RE can be composed of RLEM 3-10. 
\end{lem}
\begin{lem}\label{LEM:3-10_by_2-3_2-4}{\rm \cite{LPAM08R}}\ 
RLEM 3-10 can be composed of RLEMs 2-3 and 2-4. 
\end{lem}
\begin{lem}\label{LEM:2-3_2-4_by_3RLEM}{\rm \cite{MOAT12}}\ 
RLEMs 2-3 and 2-4 can be composed of any one of 14 non-degenerate 
3-symbol RLEMs. 
\end{lem}
By above, we obtain the next lemma that entails universality 
of all non-degenerate 3-symbol RLEMs. 
\begin{lem}\label{LEM:3RLEM_universality}{\rm \cite{MOAT12}}\ 
An RE can be constructed by any one of 14 non-degenerate 
3-symbol RLEMs. 
\end{lem}

\begin{figure}[h]
\begin{center}
 \includegraphics[scale=0.55]{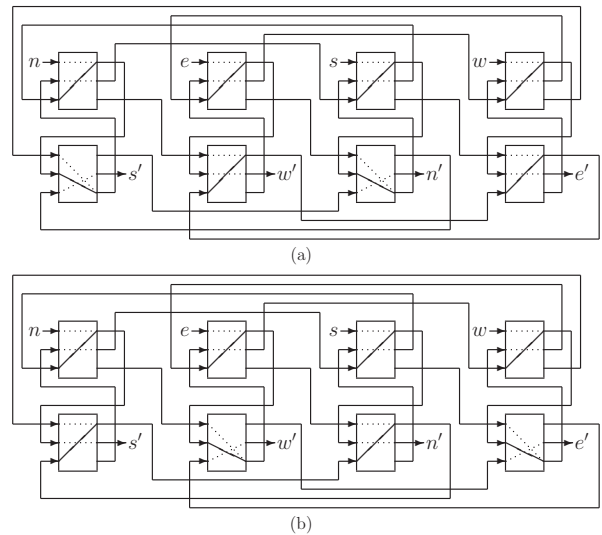}
 \caption{A circuit composed of RLEMs 3-10 that simulates RE
 \cite{MOAT12}. 
 (a) and (b) correspond to the states H and V of RE, respectively. 
 \label{FIG:REby3-10} 
 }
\end{center}
\end{figure}

\begin{figure}[h]
\begin{center}
 \includegraphics[scale=0.65]{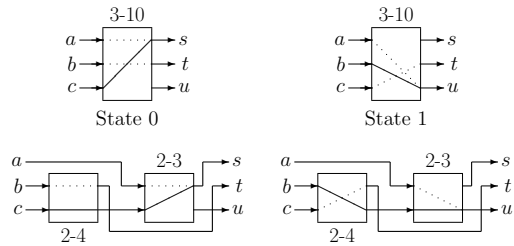}
 \caption{A circuit composed of RLEMs 2-3 and 2-4 that simulates 
 RLEM 3-10 \cite{LPAM08R}. 
 The lower left and the lower right figures correspond to 
 the states 0 and 1 of RLEM 3-10, respectively. 
 \label{FIG:3-10by2-3and4} 
 }
\end{center}
\end{figure}

\begin{figure}[t]
\begin{center}
 \includegraphics[scale=0.78]{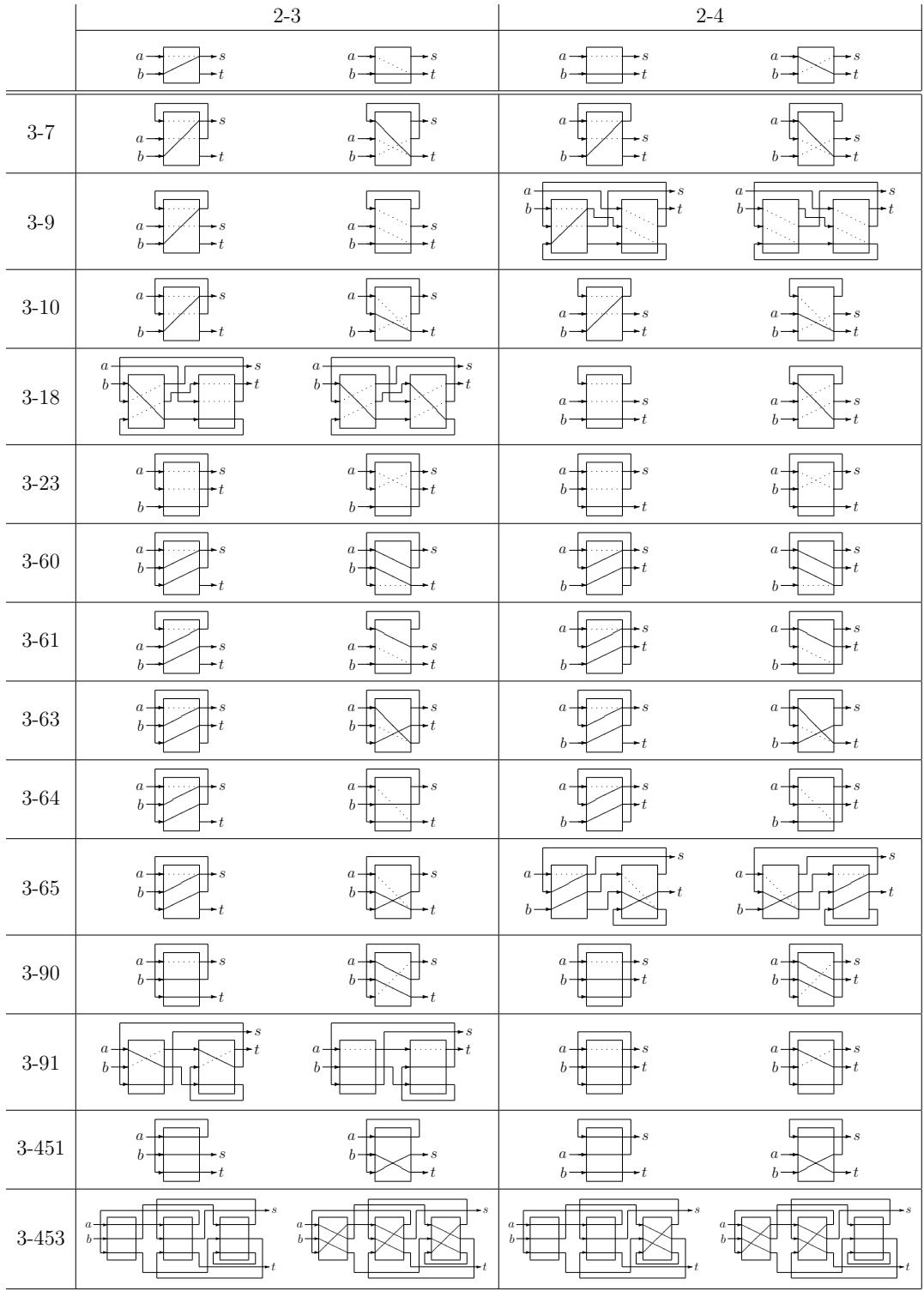}
 \caption{Circuits composed of each of 14 non-degenerate 3-symbol 
 RLEMs that simulate RLEMs 2-3 and 2-4 \cite{MOAT12}. 
 \label{FIG:2-3and4} 
 }
\end{center}
\end{figure}

Lemmas~\ref{LEM:RE_by_3-10}--\ref{LEM:2-3_2-4_by_3RLEM} are 
proved by designing circuits composed of given RLEMs 
which correctly simulate the target RLEMs.  
These circuits are shown below. 
Lemma~\ref{LEM:RE_by_3-10} is proved by a circuit 
made of RLEMs 3-10 that simulates an RE, which was first 
given in \cite{LPAM08R}.  
Later, a simpler circuit was given in \cite{MOAT12}, which is 
shown in Fig.~\ref{FIG:REby3-10}. 
Next, Lemma~\ref{LEM:3-10_by_2-3_2-4} is proved by a circuit 
made of RLEMs 2-3 and 2-4 that simulates RLEM 3-10  
shown in Fig.~\ref{FIG:3-10by2-3and4} \cite{LPAM08R}. 
Finally, Lemma~\ref{LEM:2-3_2-4_by_3RLEM} is proved 
by 28 circuits composed of each of 14 non-degenerate 
3-symbol RLEMS that simulate RLEMs 2-3 and 2-4 as 
in Fig.\ref{FIG:2-3and4}. 


\clearpage

The following lemma gives a relation between $k$-symbol RLEMs 
and $(k-1)$-symbol RLEMs. 

\begin{lem}\label{LEM:kRLEM_k-1RLEM}{\rm \cite{MOAT12}}\ 
Let $M_k$ be an arbitrary non-degenerate $k$-symbol RLEM ($k>2$). 
Then, there exists a non-degenerate $(k-1)$-symbol RLEM $M_{k-1}$ 
that can be simulated by $M_k$. 
\end{lem}

\begin{figure}[h]
\begin{center}
 \includegraphics[scale=0.6]{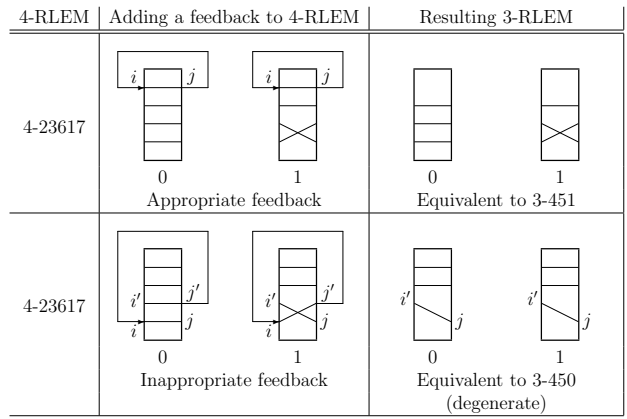}
 \caption{Making a 3-symbol RLEM by adding a feedback loop to 
 4-symbol RLEM 4-23614. 
 If the feedback is appropriate, the resulting 3-symbol RLEM 
 will be a non-degenerate one (upper row).  
 If not, it can be a degenerate one (lower row).
 \label{FIG:4RLEMto3RLEM} 
 }
\end{center}
\end{figure}

Here, we explain only a key idea of the proof of 
Lemma~\ref{LEM:kRLEM_k-1RLEM}. 
When a $k$-symbol RLEM is given, we choose one output line 
and one input line, and connect them to make a feedback loop. 
By this, we obtain a $(k-1)$-symbol RLEM. 
Fig.~\ref{FIG:4RLEMto3RLEM} shows the case of 4-symbol 
RLEM 4-23617. 
If we give an appropriate feedback loop, we can get a non-degenerate 
3-symbol RLEM (upper row of Fig.~\ref{FIG:4RLEMto3RLEM}).  
But, if we give an inappropriate feedback, then the 
resulting 3-symbol RLEM is a degenerate one (lower row 
of Fig.~\ref{FIG:4RLEMto3RLEM}).
In \cite{MOAT12}, it is proved that for a given non-degenerate 
$k$-symbol RLEM ($k>2$), we can always find a feedback loop 
by which a non-degenerate $(k-1)$-symbol RLEM can be obtained. 

By Theorem~\ref{THM:RE_universality}, and 
Lemmas~\ref{LEM:3RLEM_universality} and \ref{LEM:kRLEM_k-1RLEM}
we have the next theorem stating that almost all non-degenerate 
2-state RLEMs are universal. 
Note that universal RLEMs can simulate each other. 
\begin{thm}\label{THM:kRLEM_universality}{\rm \cite{MOAT12}}\  
Every non-degenerate 2-state $k$-symbol RLEM is universal if $k>2$. 
\end{thm}

\begin{figure}[h]
\begin{center}
 \includegraphics[scale=1.22]{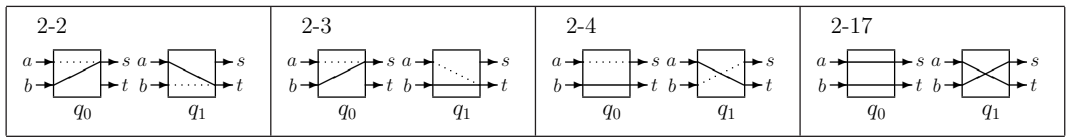}
 \caption{Four non-degenerate 2-state 2-symbol RLEMs.
 \label{FIG:nd_rlem2} 
 }
\end{center}
\end{figure}

On the other hand, there are four non-degenerate 2-state 
2-symbol RLEMs (Fig.~\ref{FIG:nd_rlem2}). 
So far, three of them have been shown to be non-universal. 

\begin{lem}\label{LEM:nonuniversality_of_2-2}{\rm \cite{MM12R}}\ 
RLEM 2-2 can simulate neither RLEM 2-3, 2-4, nor 2-17. 
\end{lem}

\begin{figure}[h]
\begin{center}
 \includegraphics[scale=1.0]{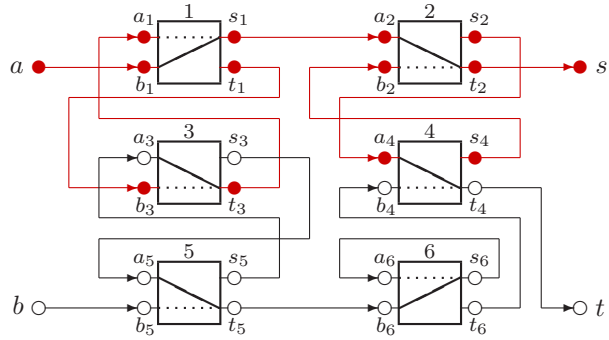}
 \caption{An example of a circuit $C$ composed of 6 copies of RLEM 2-2.
 \label{FIG:2-2_circuit} 
 }
\end{center}
\end{figure}

We give an outline of the proof of 
Lemma~\ref{LEM:nonuniversality_of_2-2}. 
Assume, on the contrary, RLEM 2-3 is simulated by a circuit $C$ 
composed of $m$ copies of RLEM 2-2 (proofs for 2-4 and 2-17 are similar). 
Let $\{a_i,b_i\}$ and $\{s_i,t_i\}$ be the sets of input and 
output ports of the $i$-th RLEM 2-2 ($i \in \{1,\ldots,m\}$) in $C$. 
Let $\{a,b\}$ and $\{s,t\}$ be those of the circuit $C$ 
(note that we assume $C$ simulates a 2-symbol RLEM), and let 
$U = \{a,b\}\,\cup\,\{s_i,t_i\ |\ i \in \{1,\ldots,m\}\}$, and 
$V = \{s,t\}\,\cup\,\{a_i,b_i\ |\ i \in \{1,\ldots,m\}\}$ 
be sets of vertices in $C$. 
The network structure of $C$ can be described by a bijection  
$f:\ U \rightarrow V$, which is called a {\em connection function}. 
In the example of Fig.~\ref{FIG:2-2_circuit}, 
$f(a)=b_1,\ f(t_1)=b_3$, etc. 
We now define a set of vertices $W$ as the smallest set that satisfies 
(i) $a \in W$, 
(ii) $x \in U \cap W \Rightarrow f(x) \in W$, 
(iii) $a_i \in V \cap W \Rightarrow s_i \in W$, and 
(iv)  $b_i \in V \cap W \Rightarrow t_i \in W$. 
Let $\overline{W} = (U \cup V)-W$. 
In Fig.~\ref{FIG:2-2_circuit}, vertices in $W$ are indicated 
by \textcolor{darkred}{\LARGE $\bullet$}, while those in 
$\overline{W}$ are by {\LARGE $\circ$}. 
Note that the set $W$ is determined only by the connection 
function $f$, not by the states of the RLEMs. 
We observe that $b \in \overline{W}$, and 
$|\{s,t\} \cap W|=1\ \wedge\ |\{s,t\} \cap \overline{W}|=1$,  
since $f$ is a bijection. 
Next, sets of RLEMs $E_{W}, E_{\overline{W}}, 
E_{W,\overline{W}} \subseteq \{1,\ldots,m\}$ are given as follows:  
$E_{W}= \{i\ |\ a_i \in W\,\wedge\,b_i \in W\}$, 
$E_{\overline{W}}= \{i\ |\ a_i \in \overline{W}\,\wedge\, 
 b_i \in \overline{W}\}$, and 
$E_{W,\overline{W}}= \{1,\ldots,m\}-(E_{W} \cup E_{\overline{W}})$. 
In Fig.~\ref{FIG:2-2_circuit}, 
$E_{W}=\{1,2\},\ E_{\overline{W}}=\{5,6\}$, and 
$E_{W,\overline{W}}=\{3,4\}$. 

Assume an input signal is given to the port $a$ or $b$ in the circuit $C$. 
Then, it visits vertices in $C$ one after another according to 
the connection function $f$ and the move function of RLEM 2-2. 
By the definitions of RLEM 2-2, $W$, $\overline{W}$, $E_{W}$, 
$E_{\overline{W}}$, and $E_{W,\overline{W}}$, we can easily see 
the following claims hold. 
\begin{enumerate}
\item 
 A signal can move from a vertex in $W$ to a vertex in $\overline{W}$, 
 or from a vertex in $\overline{W}$ to a vertex in $W$ only 
 at some RLEM in $E_{W,\overline{W}}$. 
\item 
 Assume $i \in E_{W,\overline{W}}$, and $a_i \in W$. 
 If the element $i$ is in $q_0$, then a signal at 
 $a_i \in W$ or at $b_i \in \overline{W}$  will go to $s_i \in W$. 
 Thus it is not possible to go from $W$ to $\overline{W}$ at the 
 element $i$ in $q_0$. 
 On the other hand, if the element $i$ is in $q_1$, 
 then a signal at $a_i \in W$ or at $b_i \in \overline{W}$ will 
 go to $t_i \in \overline{W}$, and thus it is not possible to go 
 from $\overline{W}$ to $W$. 
 The case $b_i \in W$ is also similar. 
\item 
 Assume $i \in E_{W,\overline{W}}$. 
 If a signal moves from a vertex in $W$ to that in $\overline{W}$, 
 or from $\overline{W}$ to $W$ at the element $i$, then 
 the element $i$ changes its state, and vice versa. 
\item 
 Let $o_1 \in \{s,t\} \cap W$ and $o_2 \in \{s,t\} \cap \overline{W}$. 
 Starting from some initial state of the circuit $C$, if a signal 
 travels from $a \in W$ to $o_2 \in \overline{W}$,  then 
 the number of times that the signal goes from $W$ to $\overline{W}$ 
 is equal to that from $\overline{W}$ to $W$ plus 1. 
 The case where a signal travels from $b \in \overline{W}$ to 
 $o_1\in W$ is also similar. 
\end{enumerate}
By above, each time a signal travels from $a \in W$ to 
$o_2 \in \overline{W}$, the number of elements in 
$E_{W,\overline{W}}$ that can make a signal move from 
$W$ to $\overline{W}$ decreases by 1. 
Similarly, each time a signal travels from $b \in \overline{W}$ 
to $o_1 \in W$, the number of elements in 
$E_{W,\overline{W}}$ that can make a signal move from 
$\overline{W}$ to $W$ decreases by 1. 
Note that, if a signal goes from $a \in W$ to $o_1 \in W$, 
or from $b \in \overline{W}$ to $o_2 \in \overline{W}$, 
the above numbers do not change. 

Consider RLEM 2-3 (see Fig.\ref{FIG:nd_rlem2}). 
Starting from $q_0$, we give an input sequence $(bb)^n$ 
($n=1,2,\ldots$) to the RLEM 2-3. 
Then, it produces an output sequence $(s\,t)^n$. 
By the assumption, the circuit $C$ composed of RLEMs 2-2 
performs this behavior. 
In either case of $s \in W \wedge\,t \in \overline{W}$ or 
$s \in \overline{W} \wedge\,t \in W$, the number of elements 
in $E_{W,\overline{W}}$ that can make a signal move from 
$\overline{W}$ to $W$ decreases indefinitely as $n$ grows large, 
since the input is always $b \in \overline{W}$. 
But, this contradicts the assumption that $C$ is composed of 
$m$ RLEMs 2-2, and thus $E_{W,\overline{W}}$ is finite. 
Hence, the circuit $C$ cannot simulate RLEM 2-3. 

For RLEM 2-4, if we give an input sequence $(ba)^n$, 
it produces $(t\,t)^n$. 
For RLEM 2-17, if we give $(bb)^n$, it produces $(s\,t)^n$.
By a similar argument as above, it is impossible for a circuit 
composed of RLEMs 2-2 to do such behaviors, and thus RLEM 2-2 
can simulate neither RLEM 2-4 nor 2-17. 
\medskip

Non-universality of RLEMs 2-3 and 2-4 is shown in \cite{MM12R}.

\begin{lem}\label{LEM:nonuniversality_of_2-3_and_2-4}{\rm \cite{MM12R}}\ 
RLEM 2-3 can simulate neither RLEM 2-4, nor 2-17, 
and RLEM 2-4 can simulate neither RLEM 2-3, nor 2-17. 
\end{lem}

By Lemmas~\ref{LEM:nonuniversality_of_2-2} and 
\ref{LEM:nonuniversality_of_2-3_and_2-4}, we have the following theorem. 

\begin{thm}\label{THM:2RLEM_non-universality}{\rm \cite{MM12R}}\ 
RLEMs 2-2, 2-3, and 2-4 are non-universal. 
\end{thm}

The following lemma says RLEM 2-2 is the weakest one 
among non-degenerate 2-state RLEMs, 

\begin{lem}\label{LEM:2-2_is_the_weakest}{\rm \cite{MM11R}}\ 
RLEM 2-2 can be simulated by any one of RLEMs 2-3, 2-4, and 2-17. 
\end{lem}

\begin{figure}[h]
\begin{center}
 \includegraphics[scale=0.53]{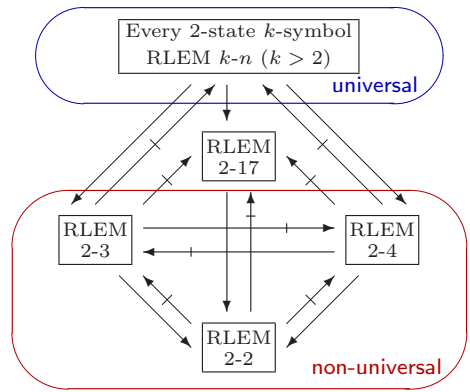}
 \caption{A hierarchy among 2-state RLEMs. 
  Here, $A \rightarrow B$ ($A \not\rightarrow B$, respectively) 
  represents that $A$ can (cannot) simulate $B$.
 \label{FIG:RLEM_hierarchy} 
 }
\end{center}
\end{figure}

Fig.~\ref{FIG:RLEM_hierarchy} summarizes the above results. 
It is not known whether RLEM 2-17 is universal or not. 
On the other hand, it is shown that any two combination among 
RLEMs 2-3, 2-4, and 2-17 is universal \cite{LPAM08R,MM11R}. 

\section{Concluding remarks}

In this survey, we discussed universality of reversible logic 
elements with memory (RLEMs), in particular 2-state RLEMs. 
It is remarkable that all non-degenerate 2-state RLEMs 
except only four are universal. 
Hence, the relation among the capability of them is rather  
simple as shown in Fig.~\ref{FIG:RLEM_hierarchy}. 
On the other hand, in the case of RLEMs with 3 or more states, 
the situation is very different. 
Even in the case of 3 states, relation among  
them seems very complex according to our partial 
experimental results \cite{MM11R}. 
In addition, we can construct many-state many-symbol non-degenerate 
RLEMs from, e.g., 2-state RLEMs 2-2.  
By this, we obtain non-universal many-state many-symbol 
non-degenerate ones, since RLEM 2-2 is non-universal. 
Thus, investigation on many-state RLEMs is left for the future study. 

\bigskip

\noindent
{\bf Acknowledgement.}\ 
This work was supported, in part, by JSPS KAKENHI Grant No.~24500017. 

%
%

\bibliographystyle{eptcs}
\bibliography{rc_morita}

\end{document}